# Relating visual attention and learning in an online instructional physics module

Razan Hamed

*Department of Physics & Astronomy, Purdue University, 525 Northwestern Ave, West Lafayette, IN, U.S.A.*

N. Sanjay Rebello, Jeremy Munsell

*Department of Physics & Astronomy, Purdue University, 525 Northwestern Ave, West Lafayette, IN, U.S.A.*

Learning using Computer-Assisted Instruction (CAI) demands a high level of attention given the tendency to be distracted and mind-wander. How does the online STEM instructor know when learners are having attentional problems and the extent to which these problems affect learning? In the present study, the visual attentional and cognitive state of physics graduate students were probed while they went through a multimedia instructional module to refresh their knowledge of Newton's II Law. Data from an eye tracker, webcam, egocentric glasses, screen recording, and mouse and keyboard events were integrated to record learners' attention overt attention to the learning environment (+/-) and thinking about learning content (+/-) to analyze students' attention spans during learning from this module. On average, learners were found to be on-task and on-screen for a vast majority of time, with evidence of mind wandering. The learning module improved the participants efficiency with which they answered the questions correctly on a post-test relative to the pre-test. Further, there is a positive albeit statistically non-significant correlation between the improvement from pre- to post-test efficiency and the time spent on-screen and on-task during the module.

# I. INTRODUCTION

Computer-Assisted Instruction (CAI) is ubiquitous and will remain so in the post COVID-19 world. A key issue with CAI is students' attention during instruction. The causal relation between attention and learning is well established. When learners ignore relevant information, or attend to irrelevant information, it reduces learning. [1–4] To address this issue, learners' attention is cued to relevant information, or remove irrelevant information, in the CAI materials. [1–4] But such solutions are only effective if learners attend to the CAI materials in the first place! Lack of sustained attention to CAI learning materials reduces learning. [5,6] However, researchers have only begun to operationalize what is meant by "students' attention span." Research has shown that a student may be looking at CAI materials on their computer screen, but not thinking about them, because they are mind-wandering. [7–9] Alternatively, a student may be looking away from their computer screen, but thinking about their learning materials, such as while note taking or reflecting. So, we must clarify the notion of attention span itself in the context of CAI environments.

# II. THEORETICAL FRAMEWORK

Research has demonstrated a way of encapsulating all these different factors that affect learning [7]. The first factor is attention: being generally inattentive vs. generally attentive to the learning environment, which can be measured by whether students are looking at the learning environment. The second factor is thinking: thinking about the learning materials vs. thinking about something else (e.g., mind wandering). The proposed research builds on D'Mello's [7] 2x2 matrix to characterize learners' attentional states during educational activities. In this framework, learners can transition between four attentional states that consider both the overt and covert aspects of attention (Fig. 1).

**Quadrant 1 (Q1)** *Top Left*: Learner visually attends to the learning environment (on-screen), while thinking about it (i.e., on-task).
**Quadrant 2 (Q2)** *Top Right*: Learner does not visually attend to the learning environment (off-screen) but thinks about it (on-task) e.g., note taking or using the calculator to solve a relevant problem.
**Quadrant 3 (Q3)** *Bottom Left*: Learner visually attends to the learning environment (on-screen) but does not think about it (off-task) e.g., mind wandering.
**Quadrant 4 (Q4)** *Bottom Right*: Learner neither visually attends to the learning environment (off-screen) nor thinks about it (off-task) e.g., distracted by a cell phone or text.

|  | Overt attention to learning materials (computer) | Overt attention *elsewhere* |
|---|---|---|
| Content-**related** thoughts | **Quadrant 1 (Q1)** **Overt** sustained attention to one or more areas of the learning materials | **Quadrant 2 (Q2)** **Covert** sustained attention (e.g., note-taking, using calculator) |
| Content-**unrelated** thoughts | **Quadrant 3 (Q3)** Covert *in*attention (mind wandering) | **Quadrant 4 (Q4)** Overt *in*attention (e.g., off-task, distracted) |

FIG. 1. The 2x2 attentional-cognition matrix.

# III. RESEARCH DESIGN

## A. Research Questions

We addressed the following research questions. RQ1: What percent of time during the instructional module did participants spend in the four attentional-cognitive states (Q1, Q2, Q3 and Q4) defined by overt attention to the learning environment (+/-), and thinking about learning content (+/-)? RQ2: How does the percentage of time that participants spent in Q1 correlate with their learning from the module?

## B. Participants

The (N=12) participants for this study were recruited from a pool of students enrolled in the physics graduate program at large U.S. midwestern land grant university. All the participants had a bachelor's degree in physics or an allied science discipline and therefore had been previously exposed, during their introductory undergraduate education, to the material covered in the instructional module. As participants had been exposed to this learning material previously in their academic preparation, this module served as a *refresher* of the learning material for the participants.

## C. Materials

The materials consisted of a pre-test, module, and post-test. The multimedia module was about 15 minutes long and was designed using the backward design strategy of [10] and consistent with [11] principles of multimedia learning. The module focused on reviewing Newton's II Law, free-body diagrams, and solving problems based on these concepts. During the module each participant was presented with a "mind-wandering" prompt in which they were explicitly asked to respond on the keyboard with a "Y" (yes) or "N" (no) to indicate whether they were mind-wandering. The mind-wandering prompts appeared randomly and continually throughout the instructional module, excluding the pre-test and post-test. The time duration between each prompt and the next was within a range of 120 seconds to 240 seconds. The pre-test and post-test each had seven multiple choice items, that included four conceptual questions and three problems that required an application of the problem-solving strategies presented in the module as well as use of scratch paper and a calculator, which were provided to the participants. In addition to the module and the tests, the study utilized annotation software which flagged instances when the participants eyes were off-screen. The software also allowed participants to view their recording of those instances in a retrospective recall interview and self-identify, with assistance from the research assistant, which attentional-cognition state (Q1, Q2, Q3, or Q4) they were in during those instances.

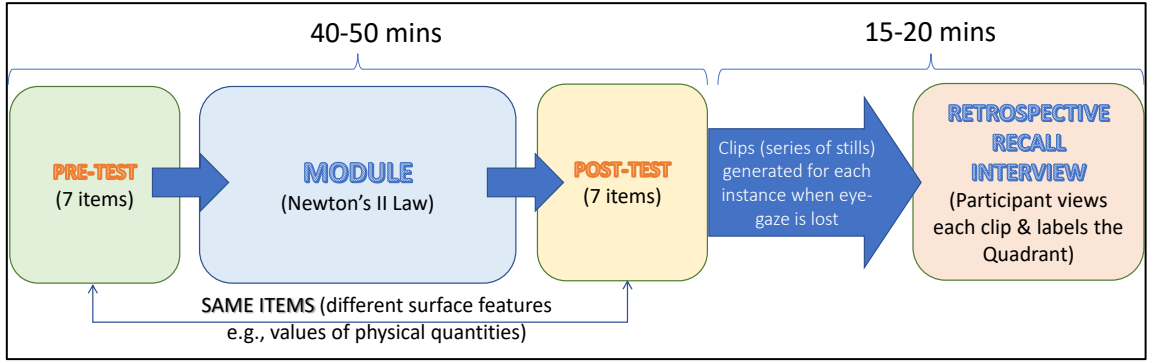

FIG. 2. Study design.

### D. Data collection process

Figure 2 shows the design of this study which took a total of 70-80 minutes for each session. During the process (pre-test, module, post-test), data from various sources (See Fig. 3) were collected: webcam, egocentric camera, screen recording, mouse and keyboard events, and eye-tracker). All sources were synchronized and recorded by specially designed software. Upon the completion of each recording, the software identified segments of time when the participant's eyes were off screen for a duration between 3-12 seconds. For each such segment of time, the software created a video clip and stored each segment as a series of 9 still frames or images.

After completion of the post-test, each participant completed a retrospective recall interview which lasted about 15-20 minutes. During this interview they were presented with video clips (each shown as a series of nine still frames) showing their screen, face (recorded from the webcam) and egocentric view (recorded from the egocentric camera). They were asked to reflect on that instant of time shown by the series of still frames and identify whether they were on/off-task and whether they were looking on/off screen (although the eye-tracker is only supposed to flag them when they are looking off-screen, sometimes it may be unable to track their eyes even when they are looking on-screen.)

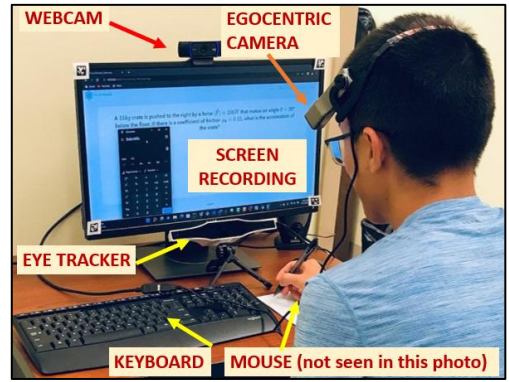

FIG. 3. Data collection sources. In the instant shown the participant is taking notes, which would be coded as off-screen/on-task (Q2).

## IV. DATA ANALYSIS

For the duration when the eye-tracker had flagged the participant as off-screen, we used data from the participants' self-reports in the retrospective recall interviews to categorize the time segment in either Q2 or Q4. For the duration when the eye-tracker had not flagged a time segment as off-screen, the participant by default was deemed to be on-screen and on-task (i.e., in Q1) unless they had responded "Y" to the mind-wandering prompt. In that case, they were deemed as mind-wandering (i.e., in Q3) for half the time between the previous prompt and the current prompt. The data from the participant self-reports was collected during the retrospective recall interview with data from participant responses to the mind-wandering prompts presented during the module.

# V. FINDINGS

## 1. Duration in Attention-Cognition States

TABLE I. Time spent by participants in each quadrant.

| Quadrant | Average Time in Quadrant |
|---|---|
| Q1 | 820.6 ± 170.7 s |
| Q2 | 63.2 ± 124.4 |
| Q3 | 72.4 ± 101.7 |
| Q4 | 2.0 ± 6.6 |

Table 1 shows the time spent by the participants in each quadrant Q1 through Q4. On average, participants spent a vast majority (85%) of their time in Q1 (on-screen and on-task). This indicates that the participants were highly engaged in the task for a vast majority of the time. This is expected given that the participants were mature graduate students who were less likely to be distracted from the task. However, we also found that on average participants were spending about 10% of their time in Q3 (on screen and off task), which indicates that they were mind wandering about 10% of the time. This too is expected given that the participants who were graduate students were familiar with the content and therefore may have found the content boring therefore were more susceptible to mind wandering than say participants such as undergraduates who were more likely to be unfamiliar with the content and may have been more interested and motivated in learning the content. Finally, we also find that participants almost a negligible percentage of their time in Q4 (off-screen and off-task), perhaps because they were completing this task in a research lab rather than in a naturalistic environment in which case off-screen and off-task behaviors might occur more frequently.

## 2. Relation between Attention-Cognition and Performance

The pre-test and post-test showed a ceiling effect, which is expected because the participants were graduate students. However, an interesting metric in this case is the efficiency in providing the correct answer to pre-test and post-test items. This metric is defined as SCORE/TIME or S/T and it is calculated for each participant on the pre-test and post-test using equation (1):

$$\frac{S}{T} = \sum_{i=1}^{M} \frac{s_i}{t_i} \quad (1)$$

where $s_i$ is the score (0 or 1) on item $i$, and $t_i$ is the time in seconds taken by the participant on item $i$, where $i$ ranges from 1 to $M$, where $M$ is the total number of items on the assessment, which in this case $M = 7$.

We argue that SCORE/TIME metric is a relevant metric for high prior knowledge participants where the pre-test and post-test scores show the ceiling effect, because this metric considers not just whether the participant answered an item correctly also it considers the time spent to answer the items. Because the time spent is in the denominator, the greater the time taken the smaller the contribution to the SCORE/TIME metric for the participant. In other words, the SCORE/TIME metric rewards participants if they answer an item correctly but penalizes participants if they take a longer time to answer the item.

We determined the gain in the score/time metric for each participant i.e., the change from pre-test to post-test, as given by equation. We wanted to test whether the change in Score/Time metric correlated with the time spent by learners in Quadrant Q1 (on-screen and on-task) as spending time in this quadrant is expected to be most conducive to learning. We found a positive, but statistically insignificant correlation coefficient (0.32) between the Gain Score/Time and the time (in seconds) spent in Quadrant Q1. (Fig. 4). Even though the result is not statistically significant, it may suggest a trend that participants who spent more time on-screen, and on-task learned more from the module, which is an expected result.

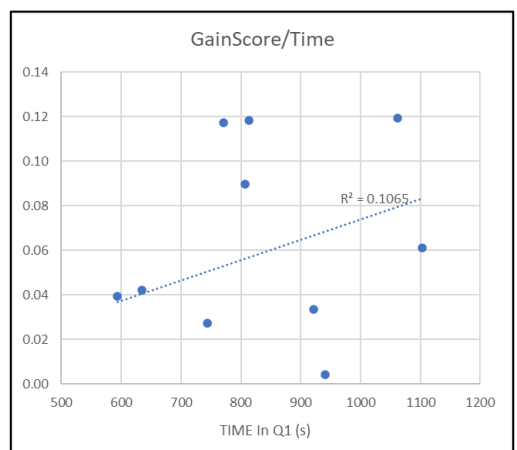

FIG. 4. Correlation between Gain Score/Time vs. Time spent in Q1.

# IV. CONCLUSIONS

To address our research questions, we find that participants spent a vast majority of their time in Q1 (on-screen and on-task), but they also spent a significant time in Q3 (on-screen and off-task) i.e., mind wandering. This could be attributed to the fact that the participants were using the module to refresh their knowledge in this area and therefore may have been bored during some aspects of the module.

A ceiling effect on the pre-test scores for these participants, given their high level of prior knowledge of the module content since they were all graduate students in physics and the module was targeted at the introductory undergraduate physics level. However, the speed with which the participants were able to answer these questions improved such that their efficiency in answering the questions correctly was improved from pre-test to post-test. Moreover, the increase in efficiency in answering the questions correctly was improved from pre-test to post-test. We also found that the increase in efficiency in answering the post-test questions correctly was positively correlated, albeit not statistically significantly, with the time they spent in Q1 i.e., on-screen and on-task.

## V. LIMITATIONS & FUTURE WORK

There are several limitations to this study that will be addressed in future work. First, the participants were high prior knowledge learners and showed a ceiling effect on the pre-test. They also were less likely to be off task and might have experienced a practice effect on the post test. We plan to repeat the study with low prior knowledge learners who might not show such an effect and show a greater propensity to be off task. Second, the study was not done in a naturalistic setting i.e., the participants were not at home or other locations where they are more likely to be distracted and mind-wander. Third, the students self-reported either being focused or mind wandering when answering the mind wandering prompts which is not a reliable way of categorizing the time that they spent looking off the screen. In future, we plan on using more objective prompts rather than using yes or no questions to detect mind wandering.

Research has shown that attention plays a critical role in online learning. However, researchers have only begun to operationally define attention or connect students' attentional states to their learning outcomes. This study is the first step to bridge the theory and methods of studying attention in cognitive science with educational practice. It points the way to making progress in understanding the connections between students' moment-to-moment attentional states and their STEM learning. It is a first step to deepen our understanding of attention/learning processes.

This study contributes to research on online learning. We have adapted and operationalized a theoretical framework to measure the moment-by-moment attention-cognition states of learners completing an online module and explore the relationship between these states and their learning outcomes on the module.

## ACKNOWLEDGMENTS

Supported in part by U.S. National Science Foundation grant DRL-2100218. We acknowledge the contributions of our collaborators who assisted with the design of the data collection software and analysis of the eye movement data.